\documentstyle[preprint,aps]{revtex}
\draft
\begin{document}
\title{{\bf Topological and Universal Aspects of Bosonized Interacting
Fermionic Systems in (2+1)d}}
\author{D.\ G.\ Barci$^\dagger$}
\address{Universidade do Estado do Rio de Janeiro,\\
Instituto de F\'\i sica, Departamento de F\'\i sica Te\'orica \\
Rua\ S\~{a}o Francisco Xavier, 524,\\
Maracan\~{a}, cep 20550, Rio de Janeiro, Brazil.}
\author{ L.\ E.\ Oxman$^{\dagger\dagger}$} 
\address{
Departamento de F\'\i sica\\
Pontif\'{\i}cia Universidade Cat\'olica\\
Cx. Postal 38071, 22452-970, Rio de Janeiro, Brazil.}
\author{S.\ P.\ Sorella$
^{\dagger\dagger\dagger}$}
\address{
Universidade do Estado do Rio de Janeiro,\\
Instituto de F\'\i sica, Departamento de F\'\i sica Te\'orica \\
Rua\ S\~{a}o Francisco Xavier, 524,\\
Maracan\~{a}, cep 20550, Rio de Janeiro, Brazil.} 
\date{November, 1998}

\maketitle

{\bf UERJ/DFT-09/98} 


\newpage

\begin{abstract}
General results on the structure of the bosonization of fermionic systems in 
$(2+1)$d are obtained. In particular, the universal character of the
bosonized topological current is established and applied to generic
fermionic current interactions. The final form of the bosonized action is
shown to be given by the sum of two terms. The first one corresponds to the
bosonization of the free fermionic action and turns out to be cast in the
form of a pure Chern-Simons term, up to a suitable nonlinear field
redefinition. We show that the second term, following from the bosonization
of the interactions, can be obtained by simply replacing the fermionic
current by the corresponding bosonized expression.
\end{abstract}

\newpage

\section{Introduction}

The method of the bosonization is a nonperturbative technique by means of
which a fermionic theory can be reformulated in terms of a bosonic one. It
has been proven to be very useful in the study of $1+1$ dimensional systems
in theoretical field theory as well as in condensed matter physics. In
particular, it displays a powerful duality mechanism which has allowed to
connect the nonperturbative (resp. perturbative) regime of the one
dimensional massive Thirring model with the perturbative (resp.
nonperturbative) one of the corresponding bosonized sine-Gordon model \cite
{Col-Man}.

Since the beginning of the nineties many efforts have been done in order to
generalize the bosonization technique to higher dimensions, with emphasis on 
$(2+1)$d fermionic systems, due to their relevance for low dimensional
condensed matter physics as, for instance, the quantum Hall effect and the
high $T_c$ superconductivity. Two different approaches have been proposed to
obtain this generalization, namely, the canonical and the functional methods.

The canonical method was employed to obtain the first results for the
bosonization of free massless fermions in $2+1$ dimensions \cite{Marino}. In
this case, the bosonization is achieved by introducing a vector gauge field.
In particular, the bosonized action turns out to be of the type of a
nonlocal Maxwell-Chern-Simons term.

While the canonical method is useful to investigate the mapping among
the fundamental fields of the fermionic and the bosonic models, the
functional method \cite{FF,Fidel,BQ1,c1,Banerjee} is appropriate to
establish the general framework for the higher dimensional bosonization
of correlation functions with insertions of the fermionic current $J^\mu
=\bar{\psi}\gamma ^\mu \psi $. \footnote{ The functional method has also
been applied to bosonize the $(1+1)$d Thirring model \cite{Naon,BQ2}.}

For instance, in the case of abelian free fermions, the generating
functional of the correlation functions of fermionic currents 
\begin{equation}
Z(s)=e^{-\Gamma (s)}=\int {\cal D}\psi {\cal D}\bar{\psi}~
e^{-K_F-i\int d^3x~\bar{\psi}s\!\!\!/\psi }\;,
\end{equation}
where 
\begin{equation}
K_F=\int d^3x~\bar{\psi}(\partial \!\!\!/+m)\psi \;,
\end{equation}
can be mapped into an equivalent bosonized form given by 
\begin{equation}
Z(s)=\int {\cal D}A_\mu ~e^{-K_B(A)-i\int d^3x~s_\mu \varepsilon ^{\mu \nu
\rho }\partial _\nu A_\rho },
\end{equation}
$s_{\mu}$ denoting the external source coupled to the spinor
current. We remark that, as already underlined, the bosonized action $K_B(A)$
is in fact given in terms of a vector field $A_\mu $ and can be cast in the
following form \cite{Fidel} 
\begin{equation}
e^{-K_B(A)}=\int {\cal D}b_\mu ~e^{-\Gamma (b)+i\int d^3x~b_\mu \varepsilon
^{\mu \nu \rho }\partial _\nu A_\rho }\;,  \label{sbosonic}
\end{equation}
where 
\begin{equation}
e^{-\Gamma (b)}=det(\partial \!\!\!/+m+ib\!\!\!/)\;.  \label{fdet}
\end{equation}
Moreover, being the bosonized action $K_B(A)$ obtained from a functional
transverse Fourier transform of the fermionic determinant, it turns out to
be gauge invariant, provided the bosonic field $A_\mu $ transforms as a
gauge field. In particular, differentiating the generating functional $Z(s)$
with respect to the external source $s_\mu $, for the Green functions of
fermionic currents one gets 
\begin{equation}
\langle J^{\mu _1}(x_1)\ldots J^{\mu _n}(x_n)\rangle _{K_F}=\langle
(\varepsilon \partial A)^{\mu _1}(x_1)\ldots (\varepsilon \partial A)^{\mu
_n}(x_n)\rangle _{K_B}~,  \label{correlators}
\end{equation}
suggesting thus the following {\it bosonization rule} for the current 
\begin{equation}
J^\mu =\bar{\psi}\gamma ^\mu \psi \leftrightarrow \varepsilon ^{\mu \nu \rho
}\partial _\nu A_\rho .  \label{current}
\end{equation}

Although the above result has been generalized in the case of the abelian
interacting massive Thirring model \cite{FF}, it remains an open question to
what extent the bosonization rule for the current (eq.\ (\ref
{current})), can be considered as a universal exact prescription for the
Green functions as well as for any fermionic interaction depending on the
current $J^\mu $. In other words, it seems natural to ask ourselves if, in
order to obtain the bosonized action for interacting fermions, we could
simply replace the fermionic current $J^\mu $ by the corresponding bosonized
topological expression $\varepsilon ^{\mu \nu \rho }\partial _\nu A_\rho $
in the interaction terms.

The aim of the present paper is to show that this important question can be
answered in the affirmative. In particular, we shall be able to establish
that, for any fermionic interaction $S^I(J)$, the bosonized form of the
generating functional $Z^{int}(s)$ can be obtained by means of the following
prescription 
\begin{equation}
K_F+i\int d^3x~s^\mu \bar{\psi}\gamma _\mu \psi +S^I(J)\leftrightarrow
K_B(A)+i\int d^3x~s^\mu \varepsilon _{\mu \nu \rho }\partial ^\nu A^\rho
+S^I(\varepsilon \partial A)\;,  \label{rule}
\end{equation}
where, remarkably enough, $K_B(A)$ is nothing but the same bosonized action
for free massive fermions of expression (\ref{sbosonic}).

From the formula (\ref{rule}) it follows that, in order to achieve the
full bosonization program, we are faced with the computation of the
exact expression of the bosonized action $K_B(A)$ for free fermions.
This amounts to compute the functional transverse Fourier transform of
the massive $(2+1)$d fermionic determinant (eq.\ \ref{fdet}).

Although a closed expression for $K_B(A)$ is still lacking, rather
interesting results have been already established. For instance, in the
infinite mass limit $m\rightarrow \infty $ the effective action
corresponding to the fermionic determinant is given exactly by a local
Chern-Simons term. Moreover, being this term quadratic, its transverse
Fourier transform can be computed straightforwardly, yielding a bosonized
action which turns out to be  a Chern-Simons term \cite{FF}. 
Another approximation
scheme was considered in ref.\cite{BFO}, where the full quadratic part of
the fermionic effective action, corresponding to the exact expression of the
vacuum polarization tensor, has been taken into account. In this case, the
approximated bosonized action takes the form of a nonlocal
Maxwell-Chern-Simons term (see also \cite{asymp}). 
It is worth underlining here that this
approximation has been proven to be very useful in order to discuss on an
equal footing both the massless and the infinite mass limit, corresponding
respectively to the bosonized actions obtained in refs.\cite{Marino,FF}.

Beyond the quadratic approximation for the fermionic determinant, the
evaluation of the bosonized action is, in general, a difficult task.
However, in a recent work \cite{Silvio}, some general features of the
fermionic determinant have been obtained without computing Feynman's
diagrams. In fact, it has been proven that the whole set of the infinite
number of one loop diagrams of the perturbative expansion of the massive
abelian determinant can be fully reset to a pure Chern-Simons term, up to a
nonlinear and nonlocal field redefinition, {\it i.e.} 
\begin{equation}
\Gamma (b)=\log \det (i\gamma ^\mu \partial _\mu +\gamma ^\mu b_\mu -m)=\eta 
{\cal S}_{CS}(\widehat{b})\;,  \label{result1}
\end{equation}
with 
\begin{equation}
{\cal S}_{CS}(\widehat{b})=\frac i2\int d^3x~\varepsilon ^{\mu \nu \rho }
\widehat{b}_\mu \partial _\nu \widehat{b}_\rho \;,  \label{result2}
\end{equation}
where $\eta $ is a numerical coefficient and $\widehat{b}_\mu =\widehat{b}
_\mu (b)$ is a gauge field obtained from the original field $b_\mu $ through
a suitable nonlinear and nonlocal redefinition.

As it will be discussed in detail in the following, the above result
(\ref {result1}) will allow us to show that the bosonized action
$K_B(A)$ of eq.(\ref{rule}) can be actually cast in the form of a pure
Chern-Simons term, up to a nonlinear and nonlocal redefinition of the
gauge field $A_\mu \rightarrow \widehat{A}_\mu (A)$. In other words, 
\begin{equation}
K_B(A)=\frac i{2\eta }\int d^3x~\varepsilon ^{\mu \nu \rho }\widehat{A}_\mu
\partial _\nu \widehat{A}_\rho \;.  \label{rule2}
\end{equation}
The two formulas (\ref{rule}) and (\ref{rule2}) represent the essence of the
present paper. While the first one exhibits the universality and the
exactness of the  bosonization rule (\ref{current}) for the current, the
second one emphasizes the general topological structure of the bosonized
action $K_B(A)$, improving then our present understanding of the whole
bosonization program in $(2+1)$d.

The present work is organized as follows. In section \S \ref{CSEA}, we
review the nonlinear field transformation leading to the Chern-Simons
structure of the fermionic effective action in $2+1$ dimensions. In section 
\S
\ref{freefer} 
we show that this topological structure also holds at
the level of the bosonized action, for any finite value of the fermionic
mass parameter. In addition, a detailed discussion of the so-called
quadratic approximation is given. Section \S \ref{Intfer} is devoted to the
study of the bosonization of a general class of interacting fermionic
systems. This is achieved by showing the validity of the universal
prescription for the bosonization of the current (\ref{rule}). A few
examples will be worked out in detail. Finally, in section \S \ref{conc}, we
collect the concluding remarks. Possible further applications will be also
outlined.

\section{Chern-Simons Structure of the Fermionic Effective Action
\label{CSEA} }

In the path integral approach to bosonization, the fermionic determinant
plays a fundamental role (cf.\ equations (\ref{sbosonic}) and (\ref{fdet})).

In $(1+1)$d the fermionic determinant for massless fermions is known to be
quadratic in the external field, and bosonization can be carried out exactly
in this case. On the other hand, the fermionic determinant for massive
fermions has not yet been computed in closed form. However, the
path-integral approach to bosonization in $(1+1)$d can bypass this fact by
exploiting the two-dimensional chiral anomaly \cite{BQ2}. The most general
external gauge field which couples to the vector fermionic current can be
decomposed as $b_\mu =\partial _\mu \eta +\varepsilon _{\mu \nu }\partial
^\nu \phi $. The component $\eta $ can be eliminated, since it corresponds
to a pure gauge. The $\phi $-term can be also thought of as a pure gauge
coupled to the $U(1)$ axial current. However, it cannot be gauged away, due
to the existence of the chiral anomaly which prevents the $(1+1)$d massless
fermionic theory from being trivial. As a consequence \cite{BQ2}, the chiral
anomalous transformation properties of the fermions in $(1+1)$d allow to
derive a set of periodicity conditions for the bosonized action which, in
the case of the massive Thirring model, lead to the well known Sine-Gordon
action.

In three dimensions a similar mechanism is not available, due to the absence
of the chiral anomaly. However, additional topological informations about
the form of the bosonized action can be derived from the general structure
of the fermionic determinant.

Indeed, in a recent work \cite{Silvio}, it has been shown that in three
dimensions an underlying topological structure does exist for the fermionic
determinant. More precisely, the resulting effective action turns out to be
of the kind of a Chern-Simons action, namely 
\begin{equation}
\Gamma (b)=\log \det (i\gamma ^\mu \partial _\mu +\gamma ^\mu b_\mu -m)=\eta 
{\cal S}_{CS}(\widehat{b})\;,  \label{res1}
\end{equation}
with 
\begin{equation}
{\cal S}_{CS}(\widehat{b})=\frac i2\int d^3x~\varepsilon ^{\mu \nu \rho }
\widehat{b}_\mu \partial _\nu \widehat{b}_\rho \;,  \label{res2}
\end{equation}
and $\widehat{b}_\mu =\widehat{b}_\mu (b)$ is obtained from the field 
$b_{\mu}$ through a nonlinear and nonlocal field redefinition.
Moreover, the redefined field $\widehat{b}_\mu $ transforms still as a
connection under gauge transformations, {\it i.e.} 
\begin{equation}
\delta b_\mu =-\partial _\mu \alpha ~,~~~~~~~\delta \widehat{b}_\mu
=-\partial _\mu \alpha \;,  \label{result3}
\end{equation}
and, of course 
\begin{equation}
\delta \Gamma (b)=\delta {\cal S}_{CS}(\widehat{b})=0\;.  \label{result4}
\end{equation}

This result is based on two well known properties of the effective
action $\Gamma (b)$: the absence of anomalies in $(2+1)$d, and the fact
that, in the infinite mass limit $m\rightarrow \infty$, the effective
action $\Gamma (b)$ takes the form of a pure local Chern-Simons term
\cite{cl}.

We shall make use extensively of these properties to extract the topological
content of the bosonized action in (2+1)d.

In order to make this paper self-contained, let us sketch here the main
lines of the reasoning that led to the result of eq.(\ref{res1}).

Let us start by considering the perturbative expansion of the effective
action, 
\begin{eqnarray}
\Gamma (b) &=&\sum_{n=2}^\infty \Gamma ^n(b)\;,  \nonumber \\
\Gamma ^n(b) &=&\int d^3x_1....d^3x_nb^{\mu _1}(x_1)...b^{\mu
_n}(x_n)<j_{\mu _1}(x_1).....j_{\mu _n}(x_n)>\;.
\end{eqnarray}
Due to the absence of anomalies, gauge invariance can be implemented at the
quantum level, {\it i.e.} 
\begin{equation}
\partial _\mu \frac{\delta \Gamma (b)}{\delta b_\mu }=0\;,
\end{equation}
implying that 
\begin{equation}
\partial _{x_1}^{\mu _1}<j_{\mu _1}(x_1).....j_{\mu _n}(x_n)>=0\;.
\label{divergence}
\end{equation}
This equation automatically decouples the longitudinal part of the gauge
field, leading to an effective action that depends only from the transverse
gauge field component $b_\mu ^T$
\begin{equation}
b_\mu ^T=(g_{\mu \nu }-\frac{\partial _\mu \partial _\nu }{\partial ^2}
)b^\nu =\frac 1{\partial ^2}\partial ^\nu F_{\mu \nu }\;.  \label{transv}
\end{equation}
Therefore, it follows that each term of the perturvative series has the form 
\begin{equation}
\Gamma ^n(b)=\int d^3y_1....d^3y_nF_{\mu _1\nu _1}(y_1)....F_{\mu _n\nu
_n}(y_n)\Omega ^{\mu _1\nu _1...\mu _n\nu _n}(y_1,...,y_n)\;,
\label{structure}
\end{equation}
with 
\begin{equation}
\Omega ^{\mu _1\nu _1...\mu _n\nu _n}=\frac 1{\left( 4\pi \right) ^n}\int
\left( \prod_{j=1}^nd^3x_j\frac{(x_j-y_j)^{\mu _j}}{\left| x_j-y_j\right| ^3}
\;\right) <j^{\nu _1}(x_1)...j^{\nu _n}(x_n)>\;.
\end{equation}
On the other hand, the two legs contribution $\Gamma ^2$ is exactly known
and can be written as 
\begin{equation}
\Gamma ^2(b)=\eta S_{CS}(b)+\int d^3xd^3y~~F_{\mu _1\nu _1}(x)\Omega ^{\mu
_1\nu _1\mu _1\mu _2}(x-y)F_{\mu _2\nu _2}(y)\;,
\end{equation}
where 
\begin{equation}
S_{CS}(b)=\frac i2\int d^3x~\varepsilon ^{\mu \nu \rho }b_\mu \partial _\nu
b_\rho ~~,  \label{lcs}
\end{equation}
is the abelian Chern-Simons action, $\Omega ^{\mu _1\nu _1\mu _1\mu _2}(x-y)$
is a well known nonlocal kernel, and $\eta $ is a regularization dependent
parameter \cite{tmym,pr,ns,red,cl}.

Thus, we can write the effective action $\Gamma (b)$ in the following form 
\begin{equation}
\Gamma (b)=\eta S_{CS}+S_\Omega \;,
\end{equation}
with 
\begin{equation}
S_\Omega =\sum_{n=2}^\infty \int d^3y_1....d^3y_nF_{\mu _1\nu
_1}(y_1)....F_{\mu _n\nu _n}(y_n)\Omega ^{\mu _1\nu _1...\mu _n\nu
_n}(y_1,...,y_n)\;.
\end{equation}

It is important to note that the first term of the fermionic effective
action is a local Chern-Simons action, while the other terms depend on the
curvature $F_{\mu \nu }$ which is nothing but the variation of the pure
Chern-Simons term 
\begin{equation}
F_{\mu \nu }=\frac 12\varepsilon _{\mu \nu \rho }\frac{\delta S_{CS}}{\delta
b_\rho }\;.
\end{equation}
This crucial feature allows to cast the fermionic effective action $\Gamma
(b)$ in the form of a Chern-Simons term evaluated at the redefined
connection $\widehat{b}_\mu $.

This statement can be formalized by means of a simple cohomological
argument. Introducing the ghost field $c$, the anti-fields $c^{*}$ and $
b_\mu ^{*}$, and the BRST nilpotent transformations 
\begin{eqnarray}
sb_\mu  &=&-\partial _\mu c\;,  \nonumber \\
sc &=&0\;,  \nonumber \\
sb_\mu ^{*} &=&\frac{\delta (\eta {\cal S}_{CS}+{\cal S}_\Omega )}{\delta
b^\mu }\;=\frac 12\varepsilon _{\mu \nu \rho }F^{\nu \rho }+\frac{\delta 
{\cal S}_\Omega (b)}{\delta b^\mu }\;,  \nonumber \\
sc^{*} &=&-\partial ^\mu b_\mu ^{*}\;,  \nonumber
\end{eqnarray}
we can show (see \cite{Silvio} for the details) that ${\cal S}_\Omega $ can
be written as an exact BRST cocycle 
\begin{equation}
{\cal S}_\Omega =s\int d^3x_1\prod_{j=2}^nd^3x_jF_{\mu _j\nu _j}(x_j)\left(
\varepsilon _{\mu _1\nu _1\rho }b^{*\rho }+\sum_{k=1}^\infty \frac{{\cal M}
_{\mu _1\nu _1}^k}{\xi ^k}\right) \Omega ^{\mu _1\nu _1..\mu _n\nu _n}\;,
\end{equation}
with ${\cal M}_{\mu _1\nu _1}^k$ being an appropriate quantity depending on
the fields, antifields and on the kernel $\Omega ^{\mu _1\nu _1..\mu _n\nu
_n}.$ This means that the term $S_\Omega $ can be reabsorved in the
Chern-Simons action by means of a nonlinear field redefinition, as it is
well known that exact cocycles correspond to pure field redefinitions (see
for instance ref.\cite{book}). Note that this cohomological argument does
not make any reference to the detailed form of the action $S_\Omega $. This
implies that any gauge invariant action containing a local Chern-Simons term 
$\eta S_{CS}$ can be rewritten, in a way similar to eq.(\ref{res1}), as a
Chern-Simons action evaluated at the corresponding redefined connection 
$\widehat{b}_\mu$, which is a nonlinear functional of the gauge field 
$b_\mu$.

In order to find an explicit formula for the nonlinear redefinition $b_\mu
\rightarrow \widehat{b}_\mu $, we can give a recursive argument by
introducing a general functional $\widehat{b}_\mu $ of the external gauge
field $b_\mu $ 
\begin{equation}
\widehat{b}_\mu =b_\mu +\sum_k\frac 1{\xi ^k}\vartheta _\mu ^k\;,
\label{achapeu}
\end{equation}
so that the following condition holds 
\begin{equation}
\eta S_{CS}(b)+\frac 1\xi \Gamma ^n(b)=\eta S_{CS}(\widehat{b})\;.
\label{CSA}
\end{equation}
The parameter $\xi $ is an arbitrary coefficient and can be set to one after
the calculations.

The computation of the coefficients $\vartheta _\mu ^k$ is straightforward.
For instance, inserting eq.(\ref{achapeu}) into eq.(\ref{CSA}) and
identifying the terms of the same order in the powers of $\xi $, for the
first two coefficients one obtains 
\begin{eqnarray}
\vartheta _\mu ^1(x) &=&\frac 14\varepsilon _{\mu \sigma \tau }\Xi ^{\sigma
\tau }\;,  \label{coeff} \\
\vartheta _\mu ^2(x) &=&-\varepsilon _{\mu \nu \rho }\int d^3y{\cal F}
_{3n}^{\sigma \tau \nu \rho }(y,x)\varepsilon _{\sigma \tau \alpha }\partial
_\beta \Xi ^{\alpha \beta }\;,  \nonumber
\end{eqnarray}
with $\Xi ^{\sigma \tau }$ and ${\cal F}_{3n}^{\sigma \tau \nu \rho }(y,x)$
given by

\begin{eqnarray}
\Xi ^{\nu \rho } &=&\frac 1\eta \int d^3x_2....d^3x_nF_{\mu _2\nu
_2}(x_2)....F_{\mu _n\nu _n}(x_n)\Omega ^{\nu \rho \mu _2\nu _2...\mu _n\nu
_n}(x,x_2,...,x_n)\;, \nonumber   \\
{\cal F}_{3n}^{\sigma \tau \nu \rho } &=&\int \left(
\prod_{j=3}^nd^3x_jF_{\mu _j\nu _j}(x_j)\right) \Omega ^{\sigma \tau \nu
\rho \mu _3\nu _3...\mu _n\nu _n}(y,x,x_3,...,x_n)\;.   \label{tt1}
\end{eqnarray}

Note that $\vartheta _\mu ^1$ and $\vartheta _\mu ^2$ are gauge invariant.
This means that $\widehat{b}_\mu $ is a gauge field, according to eqs.(\ref
{result3}) and (\ref{result4}).

With a similar procedure, it is not difficult to show that the whole
fermionic effective action $\Gamma (b)$ can be reset to a Chern-Simons
action for the gauge field $\widehat{b}_\mu $, which is a nonlinear
functional of the connection $b_\mu ,$ as shown by the eqs.(\ref{res1}) and 
(\ref{res2}). It is worth remarking here that the redefined connection 
$\widehat{b}_\mu $ is in fact $\eta $-dependent (see eq.\ (\ref{coeff})).

The coefficients $\vartheta _\mu ^k$ define an expansion of the effective
action which will be useful when discussing bosonization. A detailed
calcluation of the $\vartheta _\mu ^k$ 's can be found in ref.\cite{Silvio}.
In particular, it can be  shown that the linear component of the $\vartheta
_\mu ^k$ depends only on the vacuum polarization tensor $<J^\mu J^\nu >$, 
{\it i.e.} 
\begin{equation}
\left. \frac{\delta \vartheta _\mu ^k}{\delta b_\nu }\right| _{b=0}={\rm 
function}\;{\rm of\;}\left( <J^\mu J^\nu >\right) \;.  \label{pimunu}
\end{equation}
We can summarize this section by stating that any gauge invariant action
containing a local Chern-Simons term can be rewritten as a pure Chern-Simons
term, up to a suitable nonlinear field redefinition. In particular, the
effective action for noninteracting fermions can be interpreted in a
geometrical way as a particular Chern-Simons functional associated to the
redefined connection $\widehat{b}_\mu (b)$, depending in a nonlinear way
on the external gauge field $b_\mu $. Let us also observe that in the
particular case in which the external field $b_\mu $ is set to $b_\mu =
\widehat{b}_\mu ^{-1}(B)$, the resulting effective action will be a pure
Chern-Simons term for the gauge field $B_\mu $. In addition, as we shall see
in section \S \ref{Intfer}, changing the connection $\widehat{b}_\mu $ will
amount to turning on interactions between fermions.

In the following section we will use these properties to gain some insight
into the topological struture of the bosonization in $(2+1)$d.

\section{ 
Chern-Simons Structure of $(2+1)$d Bosonization}
{\label{freefer}}

\subsection{Free Fermions}

The aim of this subsection is to analyze the topological structure of
the bosonized action for free fermions.

In order to write a formal expression for the bosonizing action, let us
follow the path-integral approach of ref.\cite{Fidel}. We can write the
generating functional for free massive fermions as 
\begin{equation}
Z(s)=e^{-\Gamma s)}=\int {\cal D}\psi {\cal D}\bar{\psi}~e^{-\int d^3x~\bar{
\psi}(\partial \!\!\!/+m+is\!\!\!/)\psi }\;.  \label{freef}
\end{equation}
Making the following change of variables, 
\begin{eqnarray}
\psi  &\longrightarrow &e^{i\alpha (x)}\psi \;,  \nonumber \\
\bar{\psi} &\longrightarrow &\bar{\psi}e^{-i\alpha (x)}\;,
\end{eqnarray}
and noting that $Z(s)$ does not depend on $\alpha $, it follows 
\begin{equation}
Z(s)=\int {\cal D}\psi {\cal D}\bar{\psi}{\cal D}\alpha ~F(\alpha )e^{-\int
d^3x~\bar{\psi}(\partial \!\!\!/+i(s\!\!\!/+\partial \!\!\!/\alpha )+m)\psi
}\;,  \label{p-i}
\end{equation}
where $F(\alpha )$ is an arbitrary function whose effect is that of changing
the normalization factor of the generating functional. Let us introduce now
the field $b_\mu $ through the condition 
\begin{equation}
\partial _\mu \alpha =b_\mu ~~~~~~~,~~~~~~~\alpha =\frac 1{\partial
^2}\partial ^\mu b_\mu \;.
\end{equation}
In order to integrate over $b_\mu $ in the path-integral (\ref{p-i}) we must
impose that $b_\mu $ be a pure gauge. This can be done by inserting the
delta functional $\delta (\varepsilon ^{\mu \nu \rho }f_{\nu \rho }(b))$ in
the expression for $Z(s).$ Thus, 
\begin{equation}
Z(s)=\int {\cal D}\psi {\cal D}\bar{\psi}{\cal D}b_\mu ~F(\partial b)\delta
(\varepsilon ^{\mu \nu \rho }f_{\nu \rho }(b))~e^{-\int d^3x~\bar{\psi}
(\partial \!\!\!/+i(s\!\!\!/+b\!\!\!/)+m)\psi }\;.
\end{equation}
The source $s_\mu $ can be decoupled from the fermions by shifting the field 
$b_\mu $ 
\begin{equation}
b_\mu \longrightarrow b_\mu -s_\mu \;,  \label{shift}
\end{equation}
obtaining

\begin{equation}
Z(s)=\int {\cal D}b_\mu ~F(\partial b-\partial s)\delta (\varepsilon ^{\mu
\nu \rho }f_{\nu \rho }(b-s))~e^{\log \det (\partial \!\!\!/+ib\!\!\!/+m)}\;.
\end{equation}
Exponenciating the delta functional by means of a Lagrange mutiplier $A_\mu $,
we get 
\begin{equation}
Z(s)=\int {\cal D}A_\mu ~e^{-K_B(A)-i\int d^3x~\varepsilon ^{\mu \nu \rho
}A_\mu \partial _\nu s_\rho }\;,  \label{freeb}
\end{equation}
where 
\begin{equation}
e^{-K_B(A)}=\int {\cal D}b_\mu F(\partial b)~e^{-\Gamma (b)+i\int
d^3x~\varepsilon ^{\mu \nu \rho }A_\mu \partial _\nu b_\rho }\;.
\label{SbosA}
\end{equation}
Then, the exponential of the bosonizing action is obtained from the
transverse Fourier transform of the exponential of the effective action 
$\Gamma (b)$
\footnote{ We have supposed that $\partial s=0$, as $Z(s)$ is gauge
invariant and therefore depends only on the transverse component of
$s$.}. Recalling from the previous section that the fermionic effective
action $\Gamma (b)$ can be cast in the form of a Chern-Simons term (see
eq.(\ref{res1})), we have 
\begin{equation}
e^{-K_B(A)}=\int {\cal D}bF(\partial b)~e^{-\eta S_{CS}(\widehat{b})+i\int
d^3x~\varepsilon ^{\mu \nu \rho }A_\mu \partial _\nu b_\rho }\;.
\label{leff}
\end{equation}
This expression is suitable in order to obtain additional informations on
the final form of the bosonized action. Firstly, we note that the term $
K_B(A)$ is gauge invariant and depends only on the transverse component of 
$A_\mu $, 
\begin{equation}
K_B(A)=K_B(A^T)\;,
\end{equation}
because 
\begin{equation}
\int d^3x~\varepsilon ^{\mu \nu \rho }A_\mu \partial _\nu b_\rho =\int
d^3x~\varepsilon ^{\mu \nu \rho }A_\mu ^T\partial _\nu b_\rho \;.
\end{equation}
Secondly, it contains a local Chern-Simons term. In fact, taking the
infinite mass limit, we have 
\begin{equation}
\lim_{m\rightarrow \infty }e^{-K_B(A)}=\int {\cal D}b_\mu F(\partial
b)~e^{-\eta S_{CS}(b)+i\int d^3x~\varepsilon ^{\mu \nu \rho }A_\mu \partial
_\nu b_\rho }\;,  \label{lef}
\end{equation}
which yields, upon integration over $b_\mu $, the expression 
\begin{equation}
\lim_{m\rightarrow \infty }e^{-K_B(A)}=e^{-\frac 1\eta S_{CS}(A)}\;.
\label{limfree}
\end{equation}
From the above formula one can observe that the transverse Fourier transform
of the exponential of a Chern-Simons action is again the exponential of a
Chern-Simons action.

We see then that the bosonized action contains a local Chern-Simons term,
corresponding to the bosonization in the infinite mass limit first obtained
in ref.\cite{FF}. Therefore, for any finite value of the mass parameter $m$, we
can write 
\begin{equation}
K_B(A)=\frac 1\eta S_{CS}(A)+\tilde{S}_{bos}(A)\;,
\end{equation}
where 
\begin{equation}
\lim_{m\rightarrow \infty }\tilde{S}_{bos}(A)=0\;.
\end{equation}
Moreover, recalling that $\tilde{S}_{bos}(A)$ is gauge invariant, it can be
reabsorbed into the Chern-Simons action through a suitable redefinition of
the gauge connection $A_\mu $, {\it i.e.} 
\begin{equation}
K_B(A)=\frac 1\eta S_{CS}(\widehat{A})\;,
\end{equation}
where the redefined field $\widehat{A}_\mu (A)$ depends nonlinearly on 
$A_\mu $.

It is worth summarizing here the results obtained so far. We have introduced
the idea of a Chern-Simons type action $S_{CS}(\widehat{A})$, {\it i.e.}, a
Chern-Simons action for the connection $\widehat{A}_\mu (A)$, which is a
nonlinear functional of the fundamental gauge field $A_\mu $. In particular,
according to eq.(\ref{result1}), it has been proven that the fermionic
effective action has precisely this structure, namely 
\begin{equation}
\Gamma (b)=\eta S_{CS}(\widehat{b})\;.
\end{equation}
Then, using the formal bosonization rule of eq.(\ref{SbosA}) 
\begin{equation}
e^{-K_B(A)}=\int {\cal D}b_\mu F(\partial b)~e^{-\eta S_{CS}(\widehat{b}
)+i\int d^3x~\varepsilon ^{\mu \nu \rho }A_\mu \partial _\nu b_\rho }\;,
\end{equation}
we have been able to show a property that generalizes the case of infinite
mass limit  to a finite mass, stating that

{\em the transverse Fourier transform of the exponential of a Chern-Simons
type action }$S_{CS}(\widehat{b})$ {\em is again the exponential of a
Chern-Simons type action }$S_{CS}(\widehat{A})${\em , namely} 
\begin{equation}
\int {\cal D}b_\mu F(\partial b)~e^{-\eta S_{CS}(\widehat{b})-i\int
d^3x~\varepsilon ^{\mu \nu \rho }A_\mu \partial _\nu b_\rho }=e^{-\frac
1\eta S_{CS}(\widehat{A})}\;,
\end{equation}
{\em up to an appropriate field redefinition} $\widehat{A}_\mu (A).$

This is the main result of this section: the bosonized action for finite
fermionic mass is a Chern-Simons type action, {\it i.e.} 
\begin{equation}
\int d^3x~\bar{\psi}(\partial \!\!\!/+m)\psi \leftrightarrow K_B(A)=\frac
1\eta S_{CS}(\widehat{A})\;.  \label{freecase}
\end{equation}

\subsection{On the Quadratic Approximation}

In this section we exploit the quadratic approximation for the bosonization 
\cite{BFO}. We will show that an equivalent approximation is obtained by
considering the linear terms in the field transformation that allows to
reset the bosonized action into a Chern-Simons type action. Moreover, it
will become evident that this approximation leads to the {\em exact} results
only when we limit ourselves to the two point current correlation function.

From the equations (\ref{freeb}) and (\ref{freecase}) of the previous
section, we can express the fermionic partition function as 
\begin{equation}
Z(s)=\int {\cal D}A_\mu \exp \int d^3x(-\frac i{2\eta }\widehat{A}
\varepsilon \partial \widehat{A}-iA\varepsilon \partial s)\;.  \label{Zbos}
\end{equation}
On the other hand, using eq.(\ref{res1}), we can rewrite the eq.(\ref{freef}) as 
\begin{equation}
Z(s)=e^{-\Gamma (s)}=\exp (-\eta \int d^3x\frac i2\widehat{s}\varepsilon
\partial \widehat{s})=\int {\cal D}A_\mu \exp \int d^3x(-\frac i{2\eta
}A\varepsilon \partial A-iA\varepsilon \partial \widehat{s})\;.
\label{Zshat}
\end{equation}
Therefore, the current-current correlation function 
\begin{equation}
G_2=\left. \frac{\delta ^2}{\delta s\delta s}Z(s)\right| _{s=0}\;,
\label{dder}
\end{equation}
is given by 
\begin{equation}
G_2=\left[ \frac{\delta Z}{\delta \widehat{s}}\frac{\delta ^2\widehat{s}}{
\delta s\delta s}+\frac{\delta ^2Z}{\delta \widehat{s}\delta \widehat{s}}
\frac{\delta \widehat{s}}{\delta s}\frac{\delta \widehat{s}}{\delta s}
\right] _{s=0}\;.
\end{equation}
Noting that $\widehat{s}(0)=0$ and that $\frac{\delta Z}{\delta \widehat{s}}
|_0=0$\footnote{
The vacuum expectation value of the Dirac current is zero.}, it follows that 
$G_2$ depends only on the linear part $s_l(s)=\left( \left. \delta 
\widehat{s}/\delta s\right| _{s=0}\right) s$ of the redefined source 
$\widehat{s}$, 
\begin{equation}
G_2=\left. \frac{\delta ^2}{\delta s\delta s}Z_l(s)\right| _{s=0}\;,
\end{equation}
where 
\begin{equation}
Z_l(s)=\int {\cal D}A_\mu \exp \int d^3x(-\frac i{2\eta }A\varepsilon
\partial A-iA\varepsilon \partial s_l)\;.  \label{Zl1}
\end{equation}
Conversely, it can be seen from the general structure of $\widehat{s}$ given
in section \S \ref{CSEA}, that this linear part only depends on the vacuum
polarization diagram. This means that the quadratic approximation, where the
whole quadratic part of the fermionic effective action is taken into
account, is equivalent to the linear approximation $s_l(s)$ of the
transformation $\widehat{s}$.

The general form of $s_l$ can be written as 
\begin{equation}
s_l=(\alpha _{+}P_{+}+\alpha _{-}P_{-})s\;,
\end{equation}
where we have used the orthogonal projectors 
\begin{equation}
P_{\pm \mu \nu }=\frac 12(\delta _{\mu \nu }-\frac{\partial _\mu \partial
_\nu }{\partial ^2}\pm \varepsilon _{\mu \lambda \nu }\frac{\partial
_\lambda }{\sqrt{-\partial ^2}})\makebox[.5in]{,}L_{\mu \nu }=\frac{\partial
_\mu \partial _\nu }{\partial ^2}\;.  \label{projectors}
\end{equation}

It is simple to set $Z_l$ in the form of the eq.(\ref{Zbos}), {\it i.e.} 
\begin{equation}
Z_l(s)=\int {\cal D}A_\mu \exp \int d^3x(-\frac i{2\eta }A_l\varepsilon
\partial A_l-iA\varepsilon \partial s)\;,  \label{ZAl}
\end{equation}
where 
\begin{equation}
A_l(A)=(\frac 1{\alpha _{+}}P_{+}+\frac 1{\alpha _{-}}P_{-}+\frac 1\alpha
L)A\;,  \label{Al}
\end{equation}
is a linear function of $A_\mu $. This is due to the fact that $s_l$
is linear in the external source $s$. Thus, in order to find the bosonized
action in the quadratic approximation we only need to determine $\alpha _{+}$
and $\alpha _{-}$. To find them, we can rely on the fact that, using the
projectors (\ref{projectors}), the general form of the quadratic effective
action is

\begin{equation}
\frac 12\int d^3x\,s(iG(-\partial ^2)(P_{+}-P_{-})\sqrt{-\partial ^2}
+F(-\partial ^2)(P_{+}+P_{-})\partial ^2)s\;,
\end{equation}
where $F(-\partial ^2)$ and $G(-\partial ^2)$ are functions that can be
determined from the evaluation of the vacuum polarization tensor. Comparing
with 
\begin{equation}
\frac{i\eta }2\int d^3xs_l\varepsilon \partial s_l=\frac{i\eta }2\int
d^3xs_l(P_{+}-P_{-})\sqrt{-\partial ^2}s_l\;,
\end{equation}
it is straightforward to obtain 
\begin{equation}
i\eta \alpha _{+}^2=iG-F\sqrt{-\partial ^2}\makebox[.5in]{,}i\eta \alpha
_{-}^2=iG+F\sqrt{-\partial ^2}\;.  \label{alpha}
\end{equation}
Using eqs.(\ref{alpha}), (\ref{Al}) and eq.(\ref{ZAl}) we can write the
bosonized action as 
\begin{equation}
\int d^3x~A_l\varepsilon \partial A_l=2i\eta \int d^3x~A\left( P_{CS}\frac
G{F^2\partial ^2-G^2}-P_M\frac F{F^2\partial ^2-G^2}\right) A
\end{equation}
where 
\[
P_M=\frac 12(\partial ^2I-\partial \otimes \partial )=\frac
12(P_{+}+P_{-})\partial ^2\makebox[.5in]{,}P_{CS}=\frac i2\varepsilon
\partial =\frac i2(P_{+}-P_{-})\sqrt{-\partial ^2}\;, 
\]
are the ``Maxwell'' and ``Chern-Simons'' projectors, respectively.

This result coincides with that presented in ref.\cite{BFO}. In that
reference the functions $F(-\partial ^2)$ and $G(-\partial ^2)$ were
explicitly computed, obtaining for their Fourier transforms $\tilde{F}(k)$
and $\tilde{G}(k)$ 
\begin{equation}
\tilde{F}(k^2)=\frac{|m|}{4\pi k^2}\left[ 1-\frac{1-k^2/4m^2}{
(k^2/4m^2)^{1/2}}\arcsin \left( 1+\frac{4m^2}{k^2}\right) ^{-1/2}\right] \;,
\end{equation}

\begin{equation}
\tilde G(k^2)=\frac{q}{4\pi}+\frac{m}{2\pi |k|}\arcsin \left(1+\frac{4m^2}
{k^2}\right)^{-1/2}\;.
\end{equation}

These expressions have been obtained by using the Pauli-Villars
regularization. The parameter $q$ depends on the particular choice of the
regularization and, in the present case, it corresponds to the difference
between the number of positive and negative Pauli-Villars regularizing
masses. In terms of this regularization, the coefficient $\eta $ previously
introduced is given by 
\begin{equation}
\eta =\lim_{m\rightarrow \infty }\tilde{G}(k^2)=\frac 1{4\pi }(\frac
m{|m|}+q)\;.  \label{q}
\end{equation}
It is worth spending here few words about the regularization ambiguity which
appears in the expression for $\eta .$ In fact, the Pauli-Villars
regularizing masses can be introduced by means of a higher order derivative
lagrangian. Each regularizing mass is associated to a new Dirac factor with
mass $\Lambda _i$ that changes the fermion propagator so as to regularize
the ultraviolet behavior of the theory, namely 
\begin{equation}
(\partial \!\!\!/+m)^{-1}\rightarrow \left[ (\partial
\!\!\!/+m)\prod_i(\partial \!\!\!/+\Lambda _i)/\Lambda _i\right] ^{-1}\;.
\end{equation}
In $(2+1)$d, the usual Dirac lagrangian is not invariant under a parity
transformation $P$. In fact, the latter relates the two signs of the mass
parameter $m$ 
\begin{equation}
\bar{\psi}(\partial \!\!\!/+m)\psi \stackrel{P}{\rightarrow }\bar{\psi}
(\partial \!\!\!/-m)\psi \;.
\end{equation}
For this reason the fermionic effective action contains a parity breaking
Chern-Simons term whose coefficient depends on the sign of $m$: $m/|m|$.
When we introduce the Pauli-Villars regularizing masses $\Lambda _i$, we
change, in general, the parity properties of the starting 
action. Accordingly, each Pauli-Villars mass $\Lambda _i$ will yield an
additional contribution for the coefficient in front of the Chern-Simons
equal to $q_i=\Lambda _i/|\Lambda _i|$. The change of the parity properties
of the theory is measured by $q=\sum_iq_i$. Then, if we are interested in
working with an effective action having the same parity properties of the
classical fermionic field theory, we should choose an equal number of
positive and negative Pauli-Villars regularizing masses, {\it i.e.}, we have
to take $q=0$ in the eq.(\ref{q}). Note also that the Dirac factor
associated to a pair of Pauli-Villars masses $\Lambda $ and $-\Lambda $ ($q=0
$), 
\begin{equation}
(i\partial \!\!\!/+\Lambda )(i\partial \!\!\!/-\Lambda )\;,
\end{equation}
is invariant under a parity transformation, {\it i.e.}, it does not
introduce additional parity breaking terms.

Summarizing, we have seen that in order to bosonize the two point current
correlation function, all that is needed is the linear part of the nonlinear
transformation $s_\mu \rightarrow \widehat{s}_\mu $. On the other hand, this
linear part depends only on the vacuum polarization tensor, implying that
we can obtain an exact bosonization of the two point function by setting to
zero all the diagrams whose number of external legs is greater than two.
Thus, keeping the exact quadratic part of the fermionic effective action 
\cite{BFO} leads to an exact result for the two point correlation function
for any value of the fermion mass. For this reason this approximation has
led to the correct Schwinger terms in the current algebra \cite{FCL}. It is
also clear that, in order to bosonize higher order current correlation
functions, the knowledge of the full effective action is required, the
quadratic approximation being now no more sufficient.

\section{Bosonization for $(2+1)$d Interacting Fermions}
{\label{Intfer}}

In the previuos sections, we have studied the structure of the bosonization
for free fermions with arbitrary finite mass.  We shall now discuss a
general class of interacting fermionic systems whose interaction term $S^I(J)
$ depends on the spinor current $J^\mu =\bar{\psi}\gamma ^\mu \psi $.

The current correlation functions for this interacting system are determined
by the partition function $Z^{int}(s),$ where, as usual, the fermions are
probed by means of the external field $s_\mu $

\begin{eqnarray}
Z^{int}(s)=e^{-\Gamma ^{int}(s)}=\int {\cal D}\psi {\cal D}\bar{\psi}
~e^{-\int d^3x~\bar{\psi}(\partial \!\!\!/+m+is\!\!\!/)\psi -S^I(J)}\;.
\label{intef}
\end{eqnarray}
Starting with this expression, we can follow the same steps that have led
from eq.(\ref{freef}) to eqs.(\ref{freeb}), (\ref{SbosA}), to obtain 
\begin{equation}
Z^{int}(s)=\int {\cal D}A_\mu ~e^{-S_{bos}(A)-i\int d^3x~\varepsilon ^{\mu
\nu \rho }A_\mu \partial _\nu s_\rho }~,
\end{equation}
where 
\begin{equation}
e^{-S_{bos}(A)}=\int {\cal D}b_\mu ~e^{-\Gamma ^{int}(b)+i\int
d^3x~\varepsilon ^{\mu \nu \rho }A_\mu \partial _\nu b_\rho }\;,
\end{equation}
meaning that the exponential of the interacting bosonizing action is
obtained from the transverse Fourier transform of the exponential of the
interacting effective fermionic action. Although this formal structure is
interesting by its own, it turns out to be useless to improve the formal
bosonization as well as to develop a computational tool, since $\Gamma ^{int}
$ is a very complex object and so it is its transverse Fourier transform.

However, we can follow a different route  reexpressing the exponential of
the fermionic interaction $S^I(J)$ in terms of its functional Fourier
transform 
\begin{equation}
e^{-S^I(J)}=\int {\cal D}a_\mu e^{-S(a)-i\int d^3x~J_\mu a^\mu }\;.
\label{FFT}
\end{equation}
Therefore, for the generating functional of eq.(\ref{intef}) we get 
\begin{equation}
Z^{int}(s)=\int {\cal D}a_\mu {\cal D}\psi {\cal D}\bar{\psi}~e^{-\int d^3x~
\bar{\psi}(\partial \!\!\!/+m+is\!\!\!/+ia\!\!\!/)\psi -S(a)}~.  \label{x}
\end{equation}
In this way, we can interpret the fermionic interaction as an effective
interacting fermionic system minimally coupled to a dynamical quantum vector
field $a_\mu $ whose action is $S(a)$.

Note that for a generic interaction $S^I(J)$ which is not quadratic in the
spinor current $J^\mu $ the computation of $S(a)$ is, in general, not
possible. However, we will see that our final result will be given only in
terms of the known quantity $S^I(J)$, without relying on the explicit
computation of its Fourier transform $S(a).$ In other words, all that will
be needed is the  reasonable assumption that the Fourier representation (\ref
{FFT}) does in fact exist.

Performing in the expression (\ref{x}) the integration over the fermions, we
get 
\begin{equation}
Z^{int}(s)=\int {\cal D}a_\mu e^{-\Gamma (s+a)-S(a)}\;.  \label{y}
\end{equation}
where $\exp -\Gamma (s+a)=\det (\partial \!\!\!/+m+is\!\!\!/+ia\!\!\!/)$ is
the fermionic generating functional for free fermions whose bosonized form
has already been presented in eq.(\ref{freeb}). Thus, plugging eq.(\ref
{freeb}) in the eq.(\ref{y}), one has 
\begin{equation}
Z^{int}(s)=\int {\cal D}a_\mu {\cal D}A_\mu ~e^{-K_B(A)-i\int
d^3x~\varepsilon ^{\mu \nu \rho }A_\mu \partial _\nu (s_\rho +a_\rho
)-S(a)}\;.
\end{equation}
Integrating now over $a_\mu $ and using again the representation (\ref{FFT}
), we obtain finally 
\begin{equation}
Z^{int}(s)=\int {\cal D}A_\mu ~e^ { -K_B(A)-S^I(\varepsilon \partial
A)-i\int d^3x~s_\mu \varepsilon ^{\mu \nu \rho }\partial _\nu A_\rho }
\;.  \label{bosonic}
\end{equation}
As anticipated, we see that our final result relies only on the known action 
$S^I$, the action $S(a)$ being only needed at the intermediate steps.

Comparing now eq.(\ref{intef}) with eq.(\ref{bosonic}), we can read off the
bosonization rules for the kinetic fermionic action as well as for the
fermionic current. They coincide with the free case given in eqs.(\ref
{freecase}) and (\ref{current}). In particular, from eq.(\ref{bosonic}), it
follows that the bosonization rule for the fermionic interaction is 
\begin{equation}
S^I(\bar{\psi}\gamma ^\mu \psi )\leftrightarrow S^I(\varepsilon ^{\mu \nu
\rho }\partial _\nu A_\rho )\;,
\end{equation}
meaning that the equivalence between the fermionic and the bosonic current
is not only exact but is also a solid universal result, in the sense that it
is mantained in the interacting quantum theory. We can bosonize any term
involving the fermionic current replacing $J^{\mu}$ by the bosonic
topological current $\varepsilon ^{\mu \nu \rho }\partial _\nu A_\rho $.

To clarify this result, let us work out some examples. Let us first analyze
the case of a Thirring-like interaction 
\begin{equation}
Z^{int}(s)=\int {\cal D}\psi {\cal D}\bar{\psi}~e^{-\int d^3x~\bar{\psi}
(\partial \!\!\!/+m+is\!\!\!/)\psi +\frac 14(g^2)^{1-2\alpha }J_\mu \Box
^{-\alpha }J^\mu \;}\;,  \label{Tlike}
\end{equation}
where the coupling $g^2$ has the dimension of the inverse of the mass
parameter $m.$

Using the bosonization rules previously derived, we can bosonize the
expression (\ref{Tlike}) by simply replacing the fermionic current by the
corresponding bosonic topological current, {\it i.e.} 
\begin{equation}
Z^{int}(s)=\int {\cal D}A_\mu ~e^{-\frac 1\eta S_{CS}(\widehat{A})-i\int
d^3x~s_\mu \varepsilon ^{\mu \nu \rho }\partial _\nu A_\rho -\frac
14(g^2)^{1-2\alpha }F_{\mu \nu }\Box ^{-\alpha }F^{\mu \nu }}\;.
\label{Tlbos}
\end{equation}
Observe that the current-current interaction of eq.(\ref{Tlike}) can be
interpreted as coming from 
\begin{equation}
S(a)=\frac 14(g^2)^{2\alpha -1}\int d^3x~{\cal F}_{\mu \nu }(a)\Box ^{\alpha
-1}{\cal F}^{\mu \nu }(a)\;,  \label{HSG}
\end{equation}
upon integration over the field $a_\mu $. For instance, for $\alpha =0$ the
equation (\ref{Tlike}) corresponds to the local Thirring model whose
bosonization was obtained in ref.\cite{FF}. Note that, in this case, the
field $a_\mu $ in the equation (\ref{HSG}) corresponds to the usual
Hubbard-Stratonovich field, upon gauge fixing. Other values for $\alpha $
correspond to a nonlocal current-current interaction, or, equivalently, to
a generalized electrodynamics (\ref{HSG}). In particular, for $\alpha =1,$ we
recover the usual $QED_3$.

The equation (\ref{Tlbos}) shows that the Thirring-like interactions do not
modify the coefficient $\eta $ of the local Chern-Simons term. Then, as we
have discussed in section \S \ref{CSEA}, it is possible to recast the sum of
the Chern-Simons type action that bosonizes the free theory and of the term
coming from the interaction (cf.\ eq.\ (\ref{Tlbos})) in the form of a pure
Chern-Simons action with the same parameter $\eta $ ($\eta =1/(4\pi )\,m/|m|$
), {\it i.e.} 
\begin{equation}
\frac 1\eta S_{CS}(\widehat{A})\;+\frac 14(g^2)^{2\alpha -1}\int d^3x~{\cal F
}_{\mu \nu }\Box ^{\alpha -1}{\cal F}^{\mu \nu }=\frac 1\eta S_{CS}(\widehat{
A}^{int})\;,  \label{red-int}
\end{equation}
for a suitable redefinition of the gauge connection $\widehat{A}\rightarrow $
$\widehat{A}^{int}(A)$.

We see then that the bosonization formula (\ref{Tlbos}) allows to separate
the topological information of the theory from its dynamics, which is
contained in the redefinition 
$\widehat{A}^{int}=\widehat{A}^{int}(A)$. 
In particular, the eq.(\ref{red-int}) suggests the appealing idea that
a change in the connection, {\it i.e.}, $\widehat{A}\rightarrow 
\widehat{A}^{int}(A)$, corresponds to the introduction of interactions that
do not modify the coefficient $\eta$.

The understanding of the bosonization rules presented here for the
interacting case can be improved further by the following remarks on the
scaling properties displayed by the kind of fermionic interactions
considered in the eq.(\ref{Tlike}). For $\alpha \neq 1/2$ there are two
dimesionfull constants in the model, {\it i.e.}, $m$ and $g^2$. With these
constants only one dimensionless parameter can be built up, namely $\lambda
=mg^2$.

It turns out that, remarkably enough, the action of the Thirring-like model
we are dealing with depends only on the dimensionless combination $\lambda 
$.

In fact, if we choose to measure lengths (energy) in units of $1/m$ ($m$),
we can reexpress the action of the model in terms of the dimensionless
quantities 
\begin{equation}
x^{\prime }=xm\;,\mbox{\hspace{0.5 cm}and\hspace{0.5 cm}}\xi (x^{\prime
})=\frac 1m\psi (x^{\prime }/m)\;,
\end{equation}
obtaining 
\begin{equation}
S=\int d^3x^{\prime }~\bar{\xi}(i\partial \!\!\!/+1)\xi+\lambda
^{1-2\alpha }J_\mu \Box ^{-\alpha }J^\mu \;.  \label{coupling}
\end{equation}
In this case, the parameter $\lambda $ can be interpreted as a dimensionless
coupling constant. However, it cannot be considered as a coupling in a
perturbative sense, since the model is not renormalizable by power counting.
A nonperturbative technique is thus required. Therefore, the method of
bosonization could play an important role.

On the other hand, if we choose to measure lengths (energy) in units of $g^2$
($1/g^2$), we can write the action in terms of the dimensionless quantities 
\begin{equation}
x^{\prime }=x/g^2\;,\mbox{\hspace{0.5 cm}and\hspace{0.5 cm}}\phi (x^{\prime
})=g^2\psi (g^2x^{\prime })\;,
\end{equation}
yielding 
\begin{equation}
S=\int d^3x^{\prime }~\bar{\phi}(i\partial \!\!\!/+\lambda )\phi+J^\mu
\Box ^{-\alpha }J_\mu \;.  \label{mass}
\end{equation}
In this case, $\lambda $ will appear in the kinetic term instead of in the
interaction.

Of course, the expressions given in eqs.(\ref{coupling}) and (\ref{mass})
possess the same physical content, since the only diference between them is
the system of units chosen to measure lengths and energy. In particular, it
turns out that the $n\mbox{-point}$ current correlation functions written in
terms of the different variables $\psi $, $\phi $ and $\xi $ are related by 
\begin{eqnarray}
\langle J^{\mu _1}(x_1)\ldots J^{\mu _n}(x_n)\rangle _\psi &=&\frac
1{g^{2n}}\langle J^{\mu _1}(x_1/g^2)\ldots J^{\mu _n}(x_n/g^2)\rangle _\phi 
\nonumber \\
&=&m^n\langle J^{\mu _1}(mx_1)\ldots J^{\mu _n}(mx_n)\rangle _\xi \;.
\end{eqnarray}
From this equation, choosing for instance the variable $x_i=g^2\bar{x}_i$
(with $\bar{x}_i$ dimensionless) to measure lengths, we find the following
remarkable scaling relation 
\begin{equation}
\langle J^{\mu _1}(\bar{x}_1)\ldots J^{\mu _n}(\bar{x}_n)\rangle _\phi
=\lambda ^n\langle J^{\mu _1}(\lambda \bar{x}_1)\ldots J^{\mu _n}(\lambda 
\bar{x}_n)\rangle _\xi \;.
\end{equation}
Therefore, once a reference energy scale has been chosen, the parameter 
$\lambda $ determines completely the scaling behaviour of the theory. It is
worth underlining here that this interesting feature has been in fact
already observed by \cite{pr} in the computation of the effective
action of the nonabelian topological massive Yang-Mills theory.

Other models with topological properties different from those of the free
fermion case can be obtained by considering interactions that modify the
coefficient $\eta $ of the free theory.

For instance, a term of the kind 
\begin{equation}
S(a)=\theta S_{CS}(a)\;,
\end{equation}
will lead to a bosonized action of the form, 
\begin{equation}
S_{bos}=K_B+\frac 1\theta S_{CS}(A)=\frac 1\eta S_{CS}(\widehat{A})+\frac
1\theta S_{CS}(A)\;.
\end{equation}
Now, from the eq.(\ref{limfree}), we have 
\begin{equation}
\lim_{m\rightarrow \infty }S_{bos}=\frac 1{\eta ^{\prime }}S_{CS}(A)\;,
\makebox[.5in]{,}\frac 1{\eta ^{\prime }}=\frac 1\eta +\frac 1\theta \;.
\end{equation}
Thus, if $\eta ^{-1}+\theta ^{-1}\neq 0$, the resulting gauge invariant
bosonized action will contain a local Chern-Simons term (with parameter 
$1/\eta ^{\prime }$) and therefore we can rewrite it as 
\begin{equation}
S_{bos}(A)=\frac 1{\eta ^{\prime }}S_{CS}(\tilde{A}^{int})\;,
\end{equation}
for a suitable redefinition $\tilde{A}^{int}(A)$ of the gauge connection $A.$

In this case, the presence of the gauge field $\tilde{A}^{int}=\tilde{A}
^{int}(A)$ instead of the field $\widehat{A}=\widehat{A}(A)$, which
corresponds to the free case, follows from the fact that the transformation
needed to reset a gauge invariant action to a pure Chern-Simons term 
depends explicitly on the coefficient of the local Chern-Simons term (see
the discussion of section \S \ref{CSEA}). From a physical point of view,
this change implies that when a free fermion field is minimally coupled to a
Chern-Simons field not only the topological properties are modified but also
the dynamical ones. This modification is a general property of relativistic
systems. In particular, this also occurs for (quantum) {\it relativistic}
point particles due to the fact that the spin and translational degrees of
freedom are dynamically coupled \cite{Forte}.

\section{Conclusions}
{\label{conc}}

In this work general features of the bosonization in $(2+1)$d have been
analyzed. In particular, using the path integral approach, the universality
of the bosonization rule for the fermionic current has been established for
a large class of abelian fermionic interactions. In addition, we have seen
that the bosonized form of the free massive fermionic action can be written
as a pure Chern-Simons term, up to a suitable gauge field redefinition.

These two important properties can be combined together to give the
following final bosonization prescription 
\begin{equation}
K_F+i\int d^3x~s^\mu \bar{\psi}\gamma _\mu \psi +S^I(J)\leftrightarrow
K_B(A)+i\int d^3x~s^\mu \epsilon _{\mu \nu \rho }\partial ^\nu A^\rho
+S^I(\varepsilon \partial A)~,  \label{concl}
\end{equation}
where $K_B(A)$ corresponds to the bosonization of the massive free fermionic
action and is given by 
\begin{equation}
K_B(A)=\frac i{2\eta }\int d^3x~\varepsilon ^{\mu \nu \rho }\widehat{A}_\mu
\partial _\nu \widehat{A}_\rho ~.
\end{equation}
As already remarked, $\widehat{A}_\mu $ is a gauge connection depending
nonlinearly on the gauge field $A_\mu $. The above formula has to be
understood as an equivalence between the fermionic and the bosonic
representations for the generating functional of the correlators (\ref
{correlators}), upon path-integration over the corresponding fermionic or
bosonic fields.

It is remarkable that the spinor current $\bar{\psi}\gamma _\mu \psi $
can be exactly bosonized by a topologically conserved current
$\varepsilon _{\mu \nu \rho }\partial ^\nu A^\rho $ for a rather large
class of interactions $S^I(J)$. Moreover, the fact that the bosonized
fermionic kinetic term takes the form of a pure Chern-Simons action,
gives us a direct perception of the very deep topological structure
underlying the whole bosonization program in $(2+1)$d.

This structure could give a better understanding of the behaviour of
three-dimensional interacting fermionic models. This possibility relies
on the prominent role which the Chern-Simons term seems to play with
respect to any other gauge invariant term built up with the field
strenght $F_{\mu \nu } $ and its derivatives. Indeed, as proven in
refs.\cite{Silvio,lett1,lett2} , any gauge invariant action depending on
$F_{\mu \nu }$ in the presence of the Chern-Simons term can be
reabsorbed into the pure Chern-Simons action, through an appropriate
field redefinition. This peculiar feature turns out to be helpful in the
study of the low energy regime of these models, corresponding to a local
derivative expansion in the inverse of the mass parameter $m$. In this
region, the effects of the fermionic determinant can be seen indeed as
higher order perturbation terms of the pure topological Chern-Simons
action. We are left thus with the problem of perturbing a topological
field theory. It is natural therefore to expect that the physical
relevant quantities to be computed in the low energy regime are those of
the pure topological theory. In the present abelian case they are given
by the so-called loop variables of the knot theory, {\it \i .e.} 
$\int_\gamma dx^\mu A_\mu $, where $\gamma $ is a smooth closed oriented
nonintersecting curve. In other words, we expect that, in the low energy
regime, some relevant physical properties of three-dimensional fermionic
systems can be described by means of the loop correlators
\begin{equation}
\langle \int_{\gamma _1}dx^{\mu _1}A_{\mu _1}\dots \int_{\gamma _n}dx^{\mu
_n}A_{\mu _n}\rangle _{S_{bos}^{eff}(A)}~,  \label{lv}
\end{equation}
evaluated with an effective bosonized action of the type 
\begin{equation}
S_{bos}^{eff}(A)=\int d^3x~\varepsilon A\partial A+\int d^3x\left( \frac
\alpha {m^5}F^4+\frac \beta {m^9}F^6+\dots \right) ~,  \label{loop}
\end{equation}
with $\alpha $ and $\beta $ arbitrary dimensionless parameters. Needless
to say, the loop variable $\int_\gamma dx^\mu A_\mu $ corresponds, via
the Stokes theorem, to the flux of the topological current $\int_\sigma
dx^\mu \varepsilon _{\mu \nu \rho }\partial ^\nu A^\rho $ through a
surface $\sigma $ whose boundary is $\gamma $. Moreover, thanks to the
universality of the current bosonization, this amounts to study the flux
of the corresponding spinor current, {\it i.e.} $\int_\sigma ds^\mu
J_\mu $.

The analysis of the correlators (\ref{lv}) for a generic n-component
link is under investigation. Their computation, together with their
relationship with macroscopic fermionic quantities, will be reported in
a more complete and detailed paper \cite{loop}.

Finally, we hope that the identification of the topological structure of
the three-dimensional bosonization will be relevant in order to promote
this technique to a useful computational tool beyond the quadratic
approximation.

\section*{Acknowledgement}

We would like to express our gratitude to Prof.\ F.\ A.\ Schaposnik for
useful comments and discussions.

The Conselho Nacional de Pesquisa e Desenvolvimento, CNP$q$ Brazil, the
Funda\c {c}\~{a}o de Amparo \`{a} Pesquisa do Estado do Rio de Janeiro,
Faperj, and the SR2-UERJ are gratefully acknowledged for financial
support.

$^\dagger~~$ barci@dft.if.uerj.br

$^{\dagger\dagger~}$ oxman@fis.puc-rio.br

$^{\dagger\dagger\dagger}$ sorella@dft.if.uerj.br

\end{document}